\documentclass[runningheads]{llncs}
\usepackage{graphicx}
\usepackage{multirow}
\usepackage{amsmath}
\usepackage{amsfonts}
\usepackage{amssymb}
\usepackage{color}
\usepackage{graphicx}
\usepackage{url}
\usepackage{marvosym}
\usepackage{algpseudocode}

\begin{document}
\title{An Explainable Deep Framework: \\Towards Task-Specific Fusion for Multi-to-One MRI Synthesis}
\titlerunning{Explainable Task-Specific Fusion Network}
%
\author{Luyi Han\inst{1,2} \and
Tianyu Zhang\inst{1,2,3}\thanks{Tianyu Zhang and Luyi Han contributed equally to this work.} \and
Yunzhi Huang\inst{5} \and
Haoran Dou\inst{6,7} \and
Xin Wang\inst{2,3} \and
Yuan Gao\inst{2,3} \and
Chunyao Lu\inst{1,2} \and
Tao Tan\inst{2,4}\textsuperscript{(\Letter)} \and
Ritse Mann\inst{1,2}}
%
\authorrunning{L. Han et al.}
%
\institute{Department of Radiology and Nuclear Medicine, Radboud University Medical Centre, Nijmegen, The Netherlands \and
Department of Radiology, Netherlands Cancer Institute (NKI), Amsterdam, The Netherlands \and
GROW School for Oncology and Development Biology, Maastricht University, P. O. Box 616, 6200 MD, Maastricht, The Netherlands \and
Faculty of Applied Science, Macao Polytechnic University, 999078, Macao, China \and
Institute for AI in Medicine, School of Automation, Nanjing University of Information Science and Technology, Nanjing, China \and
Centre for Computational Imaging and Simulation Technologies in Biomedicine (CISTIB), School of Computing, University of Leeds, Leeds, UK \and
Biomedical Imaging Department, Leeds Institute for Cardiovascular and Metabolic Medicine (LICAMM), School of Medicine, University of Leeds, Leeds, UK
\\
\email{\{taotanjs\}@gmail.com}}
\maketitle              
\begin{abstract}
Multi-sequence MRI is valuable in clinical settings for reliable diagnosis and treatment prognosis, but some sequences may be unusable or missing for various reasons.
To address this issue, MRI synthesis is a potential solution.
Recent deep learning-based methods have achieved good performance in combining multiple available sequences for missing sequence synthesis. Despite their success, these methods lack the ability to quantify the contributions of different input sequences and estimate the quality of generated images, making it hard to be practical.
Hence, we propose an explainable task-specific synthesis network, which adapts weights automatically for specific sequence generation tasks and provides interpretability and reliability from two sides:
(1) visualize the contribution of each input sequence in the fusion stage by a trainable task-specific weighted average module;
(2) highlight the area the network tried to refine during synthesizing by a task-specific attention module.
We conduct experiments on the BraTS2021 dataset of 1251 subjects, and results on arbitrary sequence synthesis indicate that the proposed method achieves better performance than the state-of-the-art methods.
Our code is available at \url{https://github.com/fiy2W/mri_seq2seq}.

\end{abstract}
\section{Introduction}
Magnetic resonance imaging (MRI) consists of a series of pulse sequences, \textit{e.g.} T1-weighted (T1), contrast-enhanced (T1Gd), T2-weighted (T2), and T2-fluid-attenuated inversion recovery (Flair), each showing various contrast of water and fat tissues.
The intensity contrast combination of multi-sequence MRI provides clinicians with different characteristics of tissues, extensively used in disease diagnosis~\cite{mann2019breast}, lesion segmentation~\cite{menze2014multimodal}, treatment prognosis~\cite{chen2013clinical}, \textit{etc}.
However, some acquired sequences are unusable or missing in clinical settings due to incorrect machine settings, imaging artifacts, high scanning costs, time constraints, contrast agents allergies, and different acquisition protocols between hospitals~\cite{chartsias2017multimodal}.
Without rescanning or affecting the downstream pipelines, the MRI synthesis technique can generate missing sequences by leveraging redundant shared information between multiple sequences~\cite{sharma2019missing}.

Many studies have demonstrated the potential of deep learning methods for image-to-image synthesis in the field of both nature images~\cite{isola2017image,huang2018multimodal,choi2018stargan} and medical images~\cite{armanious2020medgan,uzunova2020memory,jung2021conditional}.
Most of these works introduce an autoencoder-like architecture for image-to-image translation and employ adversarial loss to generate more realistic images.
Unlike these one-to-one approaches, MRI synthesis faces the challenge of fusing complementary information from multiple input sequences.
Recent studies about multi-sequence fusion can specifically be divided into two groups: (1) image fusion and (2) feature fusion.
The image fusion approach is to concatenate sequences as a multi-channel input. Sharma \textit{et al.}~\cite{sharma2019missing} design a network with multi-channel input and output, which combines all the available sequences and reconstructs the complete sequences at once. Li \textit{et al.}~\cite{li2019diamondgan} add an availability condition branch to guide the model to adapt features for different input combinations. Dalmaz \textit{et al.}~\cite{dalmaz2022resvit} equip the synthesis model with residual transformer blocks to learn contextual features. Image-level fusion is simple and efficient but unstable -- zero-padding inputs for missing sequences lead to training unstable and slight misalignment between images can easily cause artifacts.
In contrast, efforts have been made on feature fusion, which can alleviate the discrepancy across multiple sequences, as high-level features focus on the semantic regions and are less affected by input misalignment compared to images. Zhou \textit{et al.}~\cite{zhou2020hi} design operation-based (\textit{e.g.} summation, product, maximization) fusion blocks to densely combine the hierarchical features. And Li \textit{et al.}~\cite{li2022virtual} employ self-attention modules to integrate multi-level features.
The model architectures of these methods are not flexible and difficult to adapt to various sequence combinations.
More importantly, recent studies only focus on proposing end-to-end models, lacking quantifying the contributions for different sequences and estimating the qualities of generated images.

In this work, we propose an explainable task-specific fusion sequence-to-sequence (TSF-Seq2Seq) network, which has adaptive weights for specific synthesis tasks with different input combinations and targets.
Specially, this framework can be easily extended to other tasks, such as segmentation.
Our primary contributions are as follows:
(1) We propose a flexible network to synthesize the target MRI sequence from an arbitrary combination of inputs; (2) The network shows interpretability for fusion by quantifying the contribution of each input sequence; (3) The network provides reliability for synthesis by highlighting the area the network tried to refine.

\section{Methods}
\begin{figure}[t]
    \centering
    \includegraphics[width=0.7\linewidth]{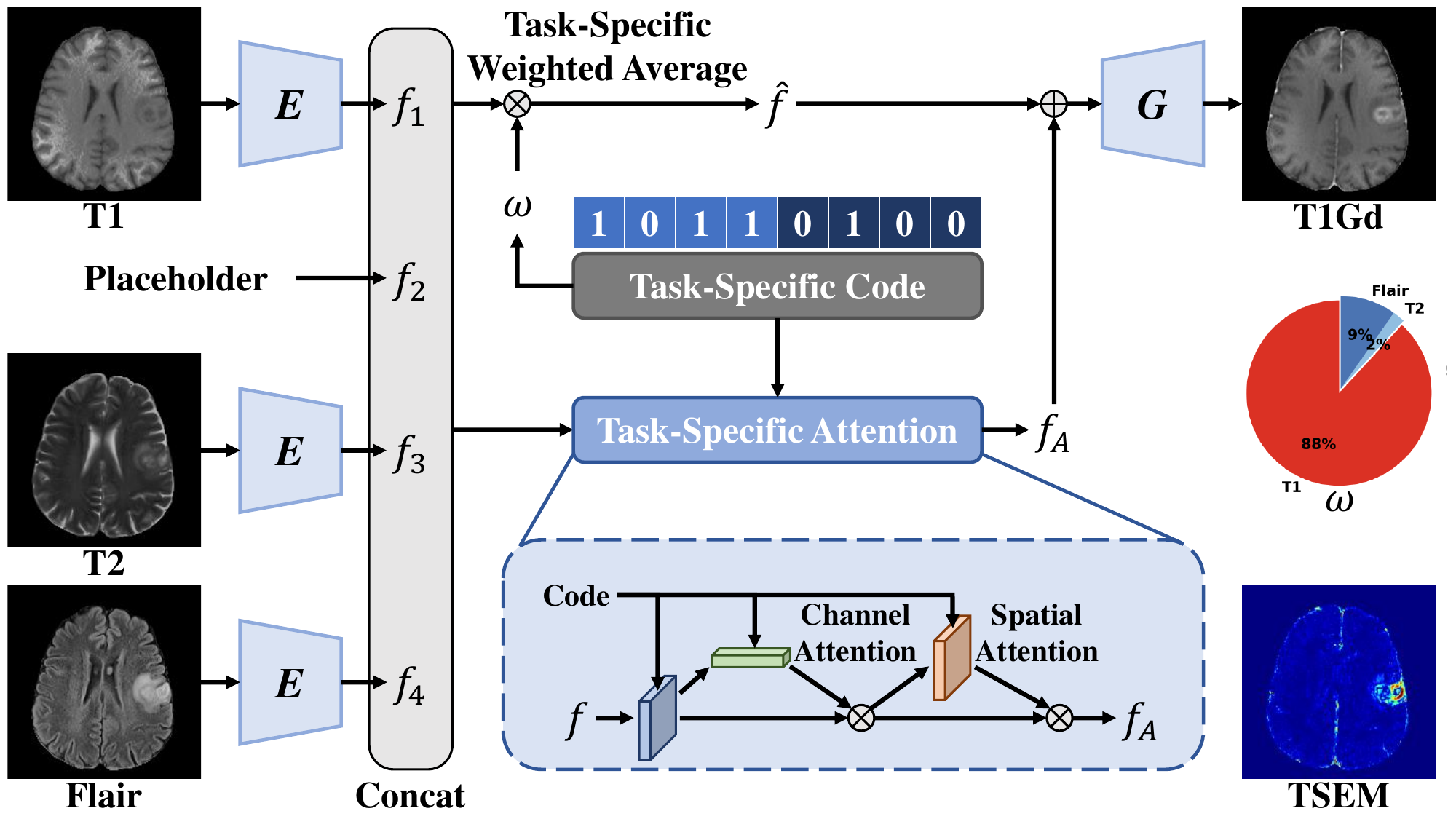}
    \caption{Overview of the TSF-Seq2Seq network. By giving the task-specific code, TSF-Seq2Seq can synthesize a target sequence from existing sequences, and meanwhile, output the weight of input sequences $\omega$ and the task-specific enhanced map (TSEM).}
    \label{fig:overview}
\end{figure}

Figure~\ref{fig:overview} illustrates the overview of the proposed TSF-Seq2Seq network.
Our network has an autoencoder-like architecture including an encoder $\mathbf{E}$, a multi-sequence fusion module, and a decoder $\mathbf{G}$.
Available MRI sequences are first encoded to features by $\mathbf{E}$, respectively. Then features from multiple input sequences are fused by giving the task-specific code, which identifies sources and targets with a binary code. Finally, the fused features are decoded to the target sequence by $\mathbf{G}$.
Furthermore, to explain the mechanism of multi-sequence fusion, our network can quantify the contributions of different input sequences and visualize the TSEM.

To leverage shared information between sequences, we use $\mathbf{E}$ and $\mathbf{G}$ from Seq2Seq~\cite{han2023synthesis}, which is a one-to-one synthetic model that integrates arbitrary sequence synthesis into single $\mathbf{E}$ and $\mathbf{G}$. They can reduce the distance between different sequences at the feature level to help more stable fusion.
Details of the multi-sequence fusion module and TSEM are described in the following sections.

\subsection{Multi-Sequence Fusion}
Define a set of $N$ sequences MRI: $\mathcal{X}=\left\{X_i|i=1,...,N \right\}$ and corresponding available indicator $\mathcal{A}\subset{\left\{1,...,N\right\}}$ and $\mathcal{A}\neq\varnothing$. Our goal is to predict the target set $\mathcal{X}_\mathcal{T}=\left\{X_i|i\notin\mathcal{A} \right\}$ by giving the available set $\mathcal{X}_\mathcal{A}=\left\{X_i|i\in\mathcal{A} \right\}$ and the corresponding task-specific code $c = \left\{ c_\text{src}, c_\text{tgt} \right\}\in\mathbb{Z}^{2N}$.
As shown in Fig.~\ref{fig:overview}, $c_\text{src}$ and $c_\text{tgt}$ are zero-one codes for the source and the target set, respectively.
To fuse multiple sequences at the feature level, we first encode images and concatenate the features as $\Vec{f} = \left\{ \mathbf{E}(X_i)|i=1,...,N \right\}$. Specifically, we use zero-filled placeholders with the same shape as $\mathbf{E}(X_i)$ to replace features of $i\notin\mathcal{A}$ to handle arbitrary input sequence combinations.
The multi-sequence fusion module includes: (1) a task-specific weighted average module for the linear combination of available features; (2) a task-specific attention module to refine the fused features.

\subsubsection{Task-Specific Weighted Average.}
The weighted average is an intuitive fusion strategy that can quantify the contribution of different sequences directly.
To learn the weight automatically, we use a trainable fully connected (FC) layer to predict the initial weight $\omega_0\in\mathbb{R}^N$ from $c$.
\begin{equation}
    \label{eq:omega0}
    \omega_0 = \text{softmax}(c\mathbf{W}+\mathbf{b})+\epsilon
\end{equation}
where $\mathbf{W}$ and $\mathbf{b}$ are weights and bias for the FC layer, $\epsilon=10^{-5}$ to avoid dividing 0 in the following equation.
To eliminate distractions and accelerate training, we force the weights of missing sequences in $\omega_0$ to be 0 and guarantee the output $\omega\in\mathbb{R}^N$ to sum to 1.
\begin{equation}
    \label{eq:omega}
    \omega = \frac{\omega_0 \cdot c_\text{src}}{\langle \omega_0, c_\text{src} \rangle}
\end{equation}
where $\cdot$ refers to the element-wise product and $\langle \cdot,\cdot \rangle$ indicates the inner product.
With the weights $\omega$, we can fuse multi-sequence features as $\hat{f}$ by the linear combination.
\begin{equation}
    \label{eq:weight}
    \hat{f} = \langle \Vec{f}, \omega \rangle
\end{equation}
Specially, $\hat{f} \equiv \mathbf{E}(X_i)$ when only one sequence $i$ is available, \textit{i.e.} $\mathcal{A}=\left\{i\right\}$.
It demonstrates that the designed $\omega$ can help the network excellently inherit the synthesis performance of pre-trained $\mathbf{E}$ and $\mathbf{G}$. In this work, we use $\omega$ to quantify the contribution of different input combinations.

\subsubsection{Task-Specific Attention.}
Apart from the sequence-level fusion of $\hat{f}$, a task-specific attention module $\mathbf{G}_A$ is introduced to refine the fused features at the pixel level.
The weights of $\mathbf{G}_A$ can adapt to the specific fusion task with the given target code.
To build a conditional attention module, we replace convolutional layers in convolutional block attention module (CBAM)~\cite{woo2018cbam} with HyperConv~\cite{han2023synthesis}.
As shown in Fig.~\ref{fig:overview}, channel attention and spatial attention can provide adaptive feature refinement guided by the task-specific code $c$ to generate residual attentional fused features $f_A$.
\begin{equation}
    \label{eq:attention}
    f_A = \mathbf{G}_A(\Vec{f}|c)
\end{equation}

\subsubsection{Loss function.}
To force both $\hat{f}$ and $\hat{f}+f_A$ can be reconstructed to the target sequence by the conditional $\mathbf{G}$, a supervised reconstruction loss is given as,
\begin{equation}
    \label{eq:reconstruction}
    \begin{aligned}
    \mathcal{L}_{rec}=&\lambda_{r}\cdot\|X'-X_\text{tgt}\|_1 + \lambda_{p}\cdot\mathcal{L}_{p}(X', X_\text{tgt}) \\
    + &\lambda_{r}\cdot\|X_A'-X_\text{tgt}\|_1 + \lambda_{p}\cdot\mathcal{L}_{p}(X_A', X_\text{tgt})
    \end{aligned}
\end{equation}
where $X'=\mathbf{G}(\hat{f}|c_\text{tgt})$, $X'_A=\mathbf{G}(\hat{f}+f_A|c_\text{tgt})$, $X_\text{tgt}\in\mathcal{X}_\mathcal{T}$, $\| \cdot \|_1$ refers to a $L_1$ loss, and $\mathcal{L}_{p}$ indicates the perceptual loss based on pre-trained VGG19. $\lambda_{r}$ and $\lambda_{p}$ are weight terms and are experimentally set to be $10$ and $0.01$.

\subsection{Task-Specific Enhanced Map}
As $f_A$ is a contextual refinement for fused features, analyzing it can help us understand more what the network tried to do.
Many studies focus on visualizing the attention maps to interpret the principle of the network, especially for the transformer modules~\cite{abnar2020quantifying,chefer2021transformer}.
However, visualization of the attention map is limited by its low resolution and rough boundary.
Thus, we proposed the TSEM by subtracting the reconstructed target sequences with and without $f_A$, which has the same resolution as the original images and clear interpretation.
\begin{equation}
    \label{eq:tsem}
    \text{TSEM} = \left| X'_A - X' \right|
\end{equation}

\section{Experiments}
\subsection{Dataset and Evaluation Metrics}
We use brain MRI images of 1,251 subjects from Brain Tumor Segmentation 2021 (BraTS2021)~\cite{baid2021rsna,bakas2017advancing,menze2014multimodal}, which includes four aligned sequences, T1, T1Gd, T2, and Flair, for each subject.
We select 830 subjects for training, 93 for validation, and 328 for testing. All the images are intensity normalized to $[-1, 1]$ and central cropped to $128\times192\times192$.
During training, for each subject, a random number of sequences are selected as inputs and the rest as targets. For validation and testing, we fixed the input combinations and the target for each subject.

\begin{table}[t]
    \centering
    \caption{Results for a set of sequences to a target sequence synthesis on BraTS2021.\label{tab:baselines}}
    \setlength{\tabcolsep}{3pt}
    \begin{tabular}{clccc}
        \hline\hline
        Number of inputs & Methods & PSNR$\uparrow$ & SSIM$\uparrow$ & LPIPS$\downarrow$ \\
        \hline
        \multirow{7}*{1}
            & Pix2Pix~\cite{isola2017image} & 25.6$\pm$3.1 & 0.819$\pm$0.086 & 15.85$\pm$9.41 \\
            & MM-GAN~\cite{sharma2019missing} & 27.3$\pm$2.4 & 0.864$\pm$0.039 & 11.47$\pm$3.76 \\
            & DiamondGAN\cite{li2019diamondgan} & 27.0$\pm$2.3 & 0.857$\pm$0.040 & 11.95$\pm$3.65 \\
            & ResViT~\cite{dalmaz2022resvit} & 26.8$\pm$2.1 & 0.857$\pm$0.037 & 11.82$\pm$3.54 \\
            \cline{2-5}
            & Seq2Seq~\cite{han2023synthesis} & 27.7$\pm$2.4 & 0.869$\pm$0.038 & 10.49$\pm$3.63  \\
            & TSF-Seq2Seq (w/o $f_A$) & 27.8$\pm$2.4 & 0.871$\pm$0.039 & \textbf{10.15$\pm$3.67} \\
            & TSF-Seq2Seq & \textbf{27.8$\pm$2.4} & \textbf{0.872$\pm$0.039} & 10.16$\pm$3.69 \\
        \hline
        \multirow{9}*{2}
            & Pix2Pix~\cite{isola2017image} (Average) & 26.2$\pm$2.7 & 0.834$\pm$0.054 & 15.84$\pm$6.05 \\
            & MM-GAN~\cite{sharma2019missing} & 28.0$\pm$2.3 & 0.878$\pm$0.037 & 10.33$\pm$3.58 \\
            & DiamondGAN\cite{li2019diamondgan} & 27.7$\pm$2.3 & 0.872$\pm$0.038 & 10.82$\pm$3.36 \\
            & ResViT~\cite{dalmaz2022resvit} & 27.7$\pm$2.2 & 0.875$\pm$0.035 & 10.53$\pm$3.26 \\
            \cline{2-5}
            & Hi-Net~\cite{zhou2020hi} & 27.1$\pm$2.3 & 0.866$\pm$0.039 & 11.11$\pm$3.76 \\
            & MMgSN-Net~\cite{li2022virtual} & 27.1$\pm$2.7 & 0.865$\pm$0.044 & 11.38$\pm$4.37 \\
            \cline{2-5}
            & Seq2Seq~\cite{han2023synthesis} (Average) & 28.2$\pm$2.2 & 0.879$\pm$0.035 & 11.11$\pm$3.72 \\
            & TSF-Seq2Seq (w/o $f_A$) & 28.0$\pm$2.4 & 0.875$\pm$0.039 & 9.89$\pm$3.63 \\
            & TSF-Seq2Seq & \textbf{28.3$\pm$2.4} & \textbf{0.882$\pm$0.038} & \textbf{9.48$\pm$3.58} \\
        \hline
        \multirow{7}*{3}
            & Pix2Pix~\cite{isola2017image} (Average) & 26.6$\pm$2.5 & 0.842$\pm$0.041 & 15.77$\pm$5.08 \\
            & MM-GAN~\cite{sharma2019missing} & 28.5$\pm$2.5 & 0.883$\pm$0.040 & 9.65$\pm$3.57 \\
            & DiamondGAN\cite{li2019diamondgan} & 28.2$\pm$2.5 & 0.877$\pm$0.041 & 10.20$\pm$3.33 \\
            & ResViT~\cite{dalmaz2022resvit} & 28.3$\pm$2.4 & 0.882$\pm$0.039 & 9.87$\pm$3.30 \\
            \cline{2-5}
            & Seq2Seq~\cite{han2023synthesis} (Average) & 28.5$\pm$2.3 & 0.880$\pm$0.038 & 11.61$\pm$3.87 \\
            & TSF-Seq2Seq (w/o $f_A$) & 28.3$\pm$2.6 & 0.876$\pm$0.044 & 9.61$\pm$4.00 \\
            & TSF-Seq2Seq & \textbf{28.8$\pm$2.6} & \textbf{0.887$\pm$0.042} & \textbf{8.89$\pm$3.80} \\
        \hline\hline
    \end{tabular}
\end{table}

\begin{figure}
    \begin{minipage}{0.19\linewidth}
        \centerline{\includegraphics[width=\textwidth]{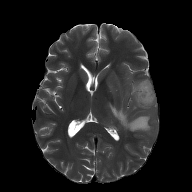}}
    \end{minipage}
    \begin{minipage}{0.19\linewidth}
        \centerline{\includegraphics[width=\textwidth]{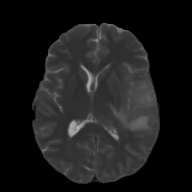}}
    \end{minipage}
    \begin{minipage}{0.19\linewidth}
        \centerline{\includegraphics[width=\textwidth]{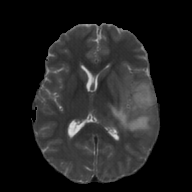}}
    \end{minipage}
    \begin{minipage}{0.19\linewidth}
        \centerline{\includegraphics[width=\textwidth]{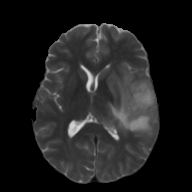}}
    \end{minipage}
    \begin{minipage}{0.19\linewidth}
        \centerline{\includegraphics[width=\textwidth]{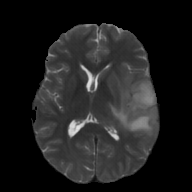}}
    \end{minipage}

    \vspace{2pt}

    \begin{minipage}{0.19\linewidth}
        \centerline{Target}
    \end{minipage}
    \begin{minipage}{0.19\linewidth}
       \centerline{Pix2Pix}
    \end{minipage}
    \begin{minipage}{0.19\linewidth}
        \centerline{MM-GAN}
    \end{minipage}
    \begin{minipage}{0.19\linewidth}
        \centerline{DiamondGAN}
    \end{minipage}
    \begin{minipage}{0.19\linewidth}
        \centerline{ResViT}
    \end{minipage}

    \vspace{5pt}
    
    \begin{minipage}{0.19\linewidth}
        \centerline{\includegraphics[width=\textwidth]{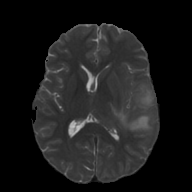}}
    \end{minipage}
    \begin{minipage}{0.19\linewidth}
        \centerline{\includegraphics[width=\textwidth]{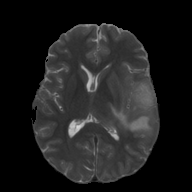}}
    \end{minipage}
    \begin{minipage}{0.19\linewidth}
        \centerline{\includegraphics[width=\textwidth]{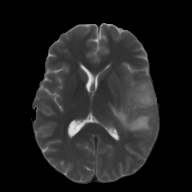}}
    \end{minipage}
    \begin{minipage}{0.19\linewidth}
        \centerline{\includegraphics[width=\textwidth]{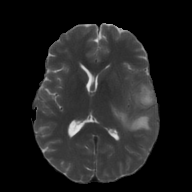}}
    \end{minipage}
    \begin{minipage}{0.19\linewidth}
        \centerline{\includegraphics[width=\textwidth]{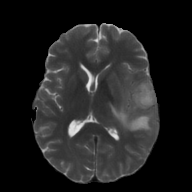}}
    \end{minipage}

    \vspace{2pt}

    \begin{minipage}{0.19\linewidth}
        \centerline{Hi-Net}
    \end{minipage}
    \begin{minipage}{0.19\linewidth}
        \centerline{MMgSN-Net}
    \end{minipage}
    \begin{minipage}{0.19\linewidth}
        \centerline{Seq2Seq}
    \end{minipage}
    \begin{minipage}{0.19\linewidth}
        \centerline{w/o $f_A$}
    \end{minipage}
    \begin{minipage}{0.19\linewidth}
        \centerline{Proposed}
    \end{minipage}

	\caption{Examples of synthetic T2 of comparison methods given the combination of T1Gd and Flair.} \label{fig:baselines}
\end{figure}

The synthesis performance is quantified using the metrics of peak signal noise rate (PSNR), structural similarity index measure (SSIM), and learned perceptual image patch similarity (LPIPS)~\cite{zhang2018unreasonable}, which evaluate from intensity, structure, and perceptual aspects.

\subsection{Implementation Details}
The models are implemented with PyTorch and trained on the NVIDIA GeForce RTX 3090 Ti GPU.
The $\mathbf{E}$ and $\mathbf{G}$ from Seq2Seq are pre-trained using the Adam optimizer with an initial learning rate of $2\times 10^{-4}$ and a batch size of 1 for 1,000,000 steps, taking about 60 hours. Then we finetune the TSF-Seq2Seq with the frozen $\mathbf{E}$ using the Adam optimizer with an initial learning rate of $10^{-4}$ and a batch size of 1 for another 300,000 steps, taking about 40 hours.

\subsection{Quantitative Results}
We compare our method with one-to-one translation, image-level fusion, and feature-level fusion methods. One-to-one translation methods include Pix2Pix~\cite{isola2017image} and Seq2Seq~\cite{han2023synthesis}. Image-level fusion methods consist of MM-GAN~\cite{sharma2019missing}, DiamondGAN~\cite{li2019diamondgan}, and ResViT~\cite{dalmaz2022resvit}. Feature-level fusion methods include Hi-Net~\cite{zhou2020hi} and MMgSN-Net~\cite{li2022virtual}.
Figure~\ref{fig:baselines} shows the examples of synthetic T2 of comparison methods input with the combinations of T1Gd and Flair.
Table~\ref{tab:baselines} reports the sequence synthesis performance for comparison methods organized by the different numbers of input combinations.
Note that, for multiple inputs, one-to-one translation methods synthesize multiple outputs separately and average them as one. And Hi-Net~\cite{zhou2020hi} and MMgSN-Net~\cite{li2022virtual} only test on the subset with two inputs due to fixed network architectures.
As shown in Table~\ref{tab:baselines}, the proposed method achieves the best performance in different input combinations.

\subsection{Ablation Study}
We compare two components of our method, including (1) task-specific weighted average and (2) task-specific attention, by conducting an ablation study between Seq2Seq, TSF-Seq2Seq (w/o $f_A$), and TSF-Seq2Seq.
TSF-Seq2Seq (w/o $f_A$) refers to the model removing the task-specific attention module.
As shown in Table~\ref{tab:baselines}, when only one sequence is available, our method can inherit the performance of Seq2Seq and achieve slight improvements.
For multi-input situations, the task-specific weighted average can decrease LPIPS to achieve better perceptual performance. And task-specific attention can refine the fused features to achieve the best synthesis results.

\begin{figure}[t]
    \begin{minipage}{0.49\linewidth}
        \centerline{\includegraphics[width=\textwidth]{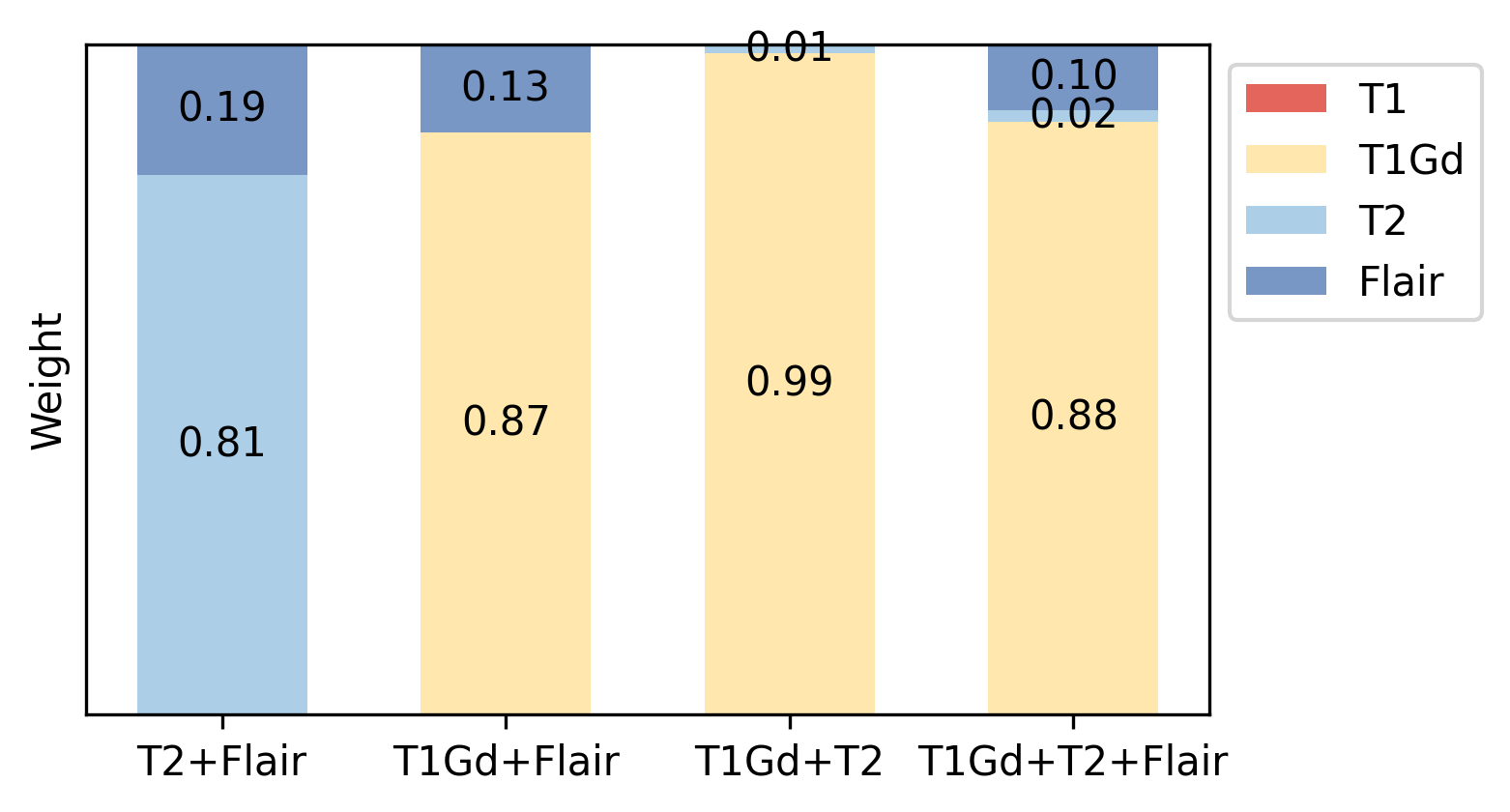}}
    \end{minipage}
    \begin{minipage}{0.49\linewidth}
        \centerline{\includegraphics[width=\textwidth]{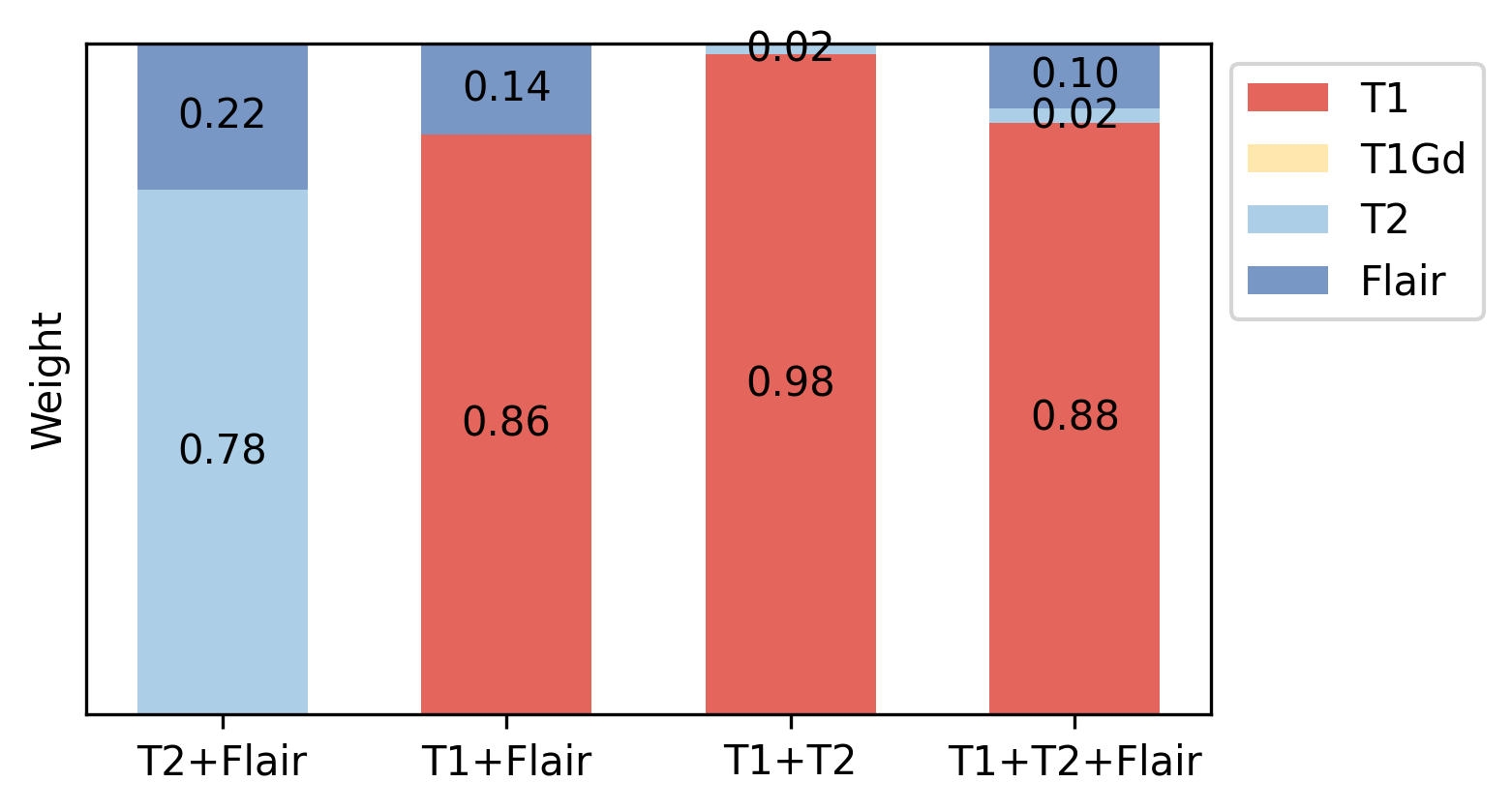}}
    \end{minipage}


    \begin{minipage}{0.49\linewidth}
        \centerline{(a)}
    \end{minipage}
    \begin{minipage}{0.49\linewidth}
        \centerline{(b)}
    \end{minipage}
    
    
    \begin{minipage}{0.49\linewidth}
        \centerline{\includegraphics[width=\textwidth]{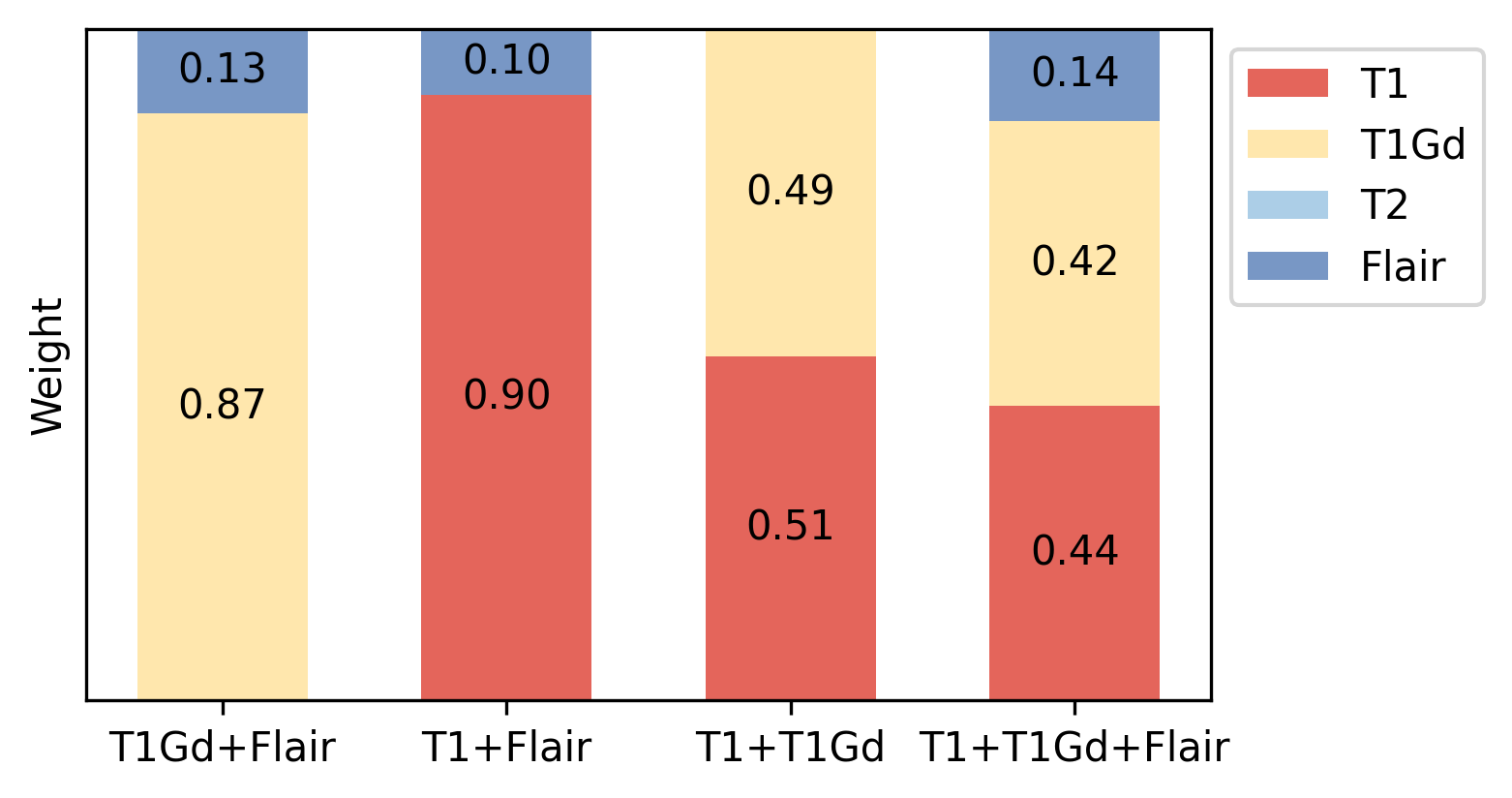}}
    \end{minipage}
    \begin{minipage}{0.49\linewidth}
        \centerline{\includegraphics[width=\textwidth]{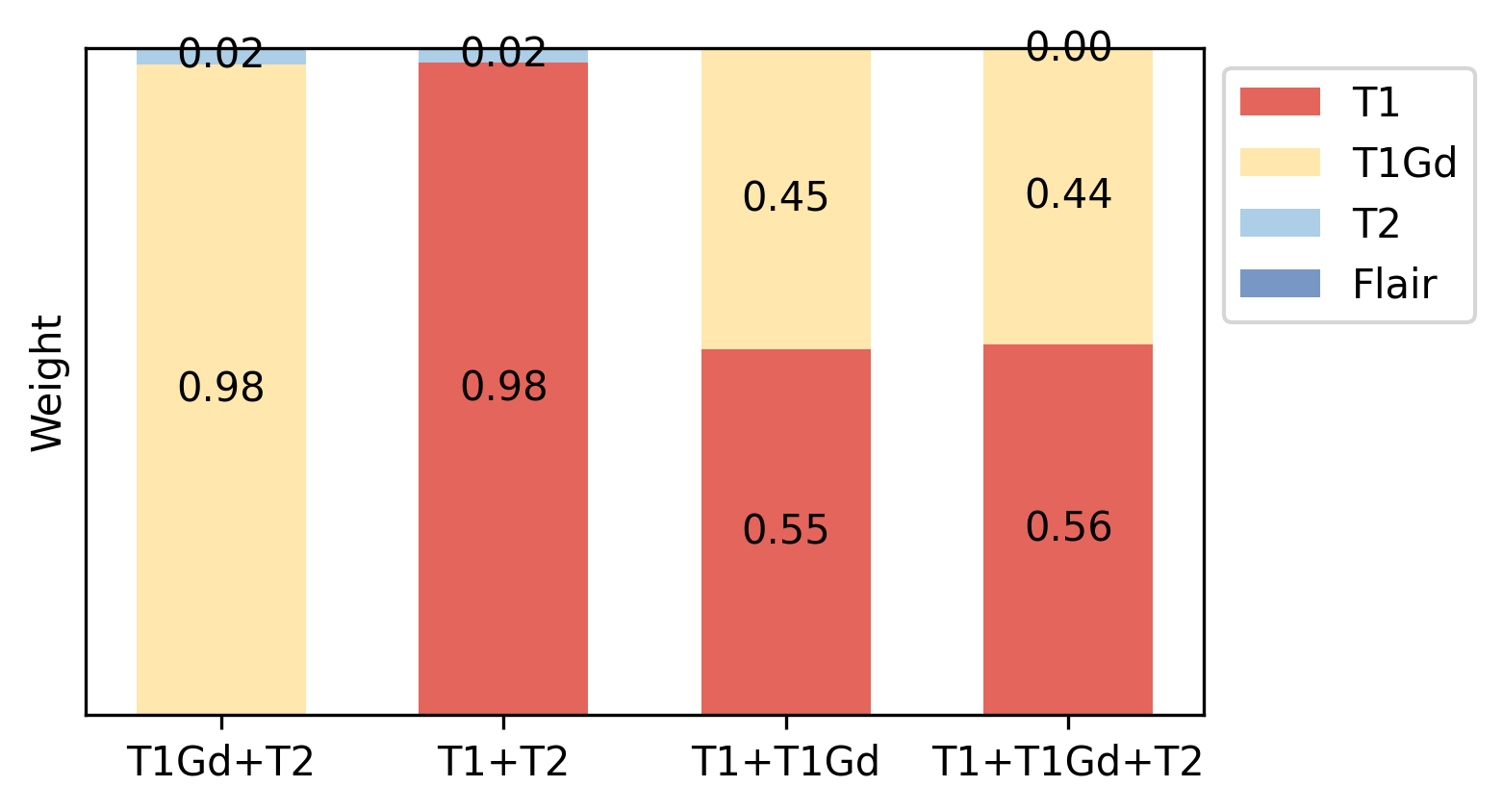}}
    \end{minipage}


    \begin{minipage}{0.49\linewidth}
        \centerline{(c)}
    \end{minipage}
    \begin{minipage}{0.49\linewidth}
        \centerline{(d)}
    \end{minipage}
	
	\caption{Bar chart for the weights of the input set of sequences to synthesize different target sequences: (a) T1; (b) T1Gd; (c) T2; (d) Flair.} \label{fig:weight}
\end{figure}

\subsection{Interpretability Visualization}
The proposed method not only achieves superior synthesis performance but also has good interpretability. In this section, we will visualize the contribution of different input combinations and TSEM.

\subsubsection{Sequence Contribution.}
We use $\omega$ in Eq.~\ref{eq:omega} to quantify the contribution of different input combinations for synthesizing different target sequences. Figure~\ref{fig:weight} shows the bar chart for the sequence contribution weight $\omega$ with different task-specific code $c$. As shown in Fig.~\ref{fig:weight}, both T1 and T1Gd contribute greatly to the sequence synthesis of each other, which is expected because T1Gd are T1-weighted scanning after contrast agent injection, and the enhancement between these two sequences is indispensable for cancer detection and diagnosis.
The less contribution of T2, when combined with T1 and/or T1Gd, is consistent with the clinical findings~\cite{zhou2020hi,zhang2023important} that T2 can be well-synthesized by T1 and/or T1Gd.

\begin{figure}[t]
    \begin{minipage}{0.12\linewidth}
        \centerline{Target}
    \end{minipage}
    \begin{minipage}{0.2\linewidth}
        \centerline{\includegraphics[width=\textwidth]{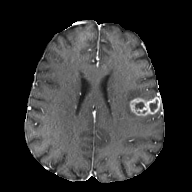}}
    \end{minipage}
    \begin{minipage}{0.2\linewidth}
        \centerline{\includegraphics[width=\textwidth]{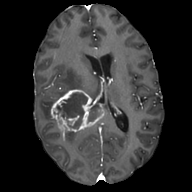}}
    \end{minipage}
    \begin{minipage}{0.2\linewidth}
        \centerline{\includegraphics[width=\textwidth]{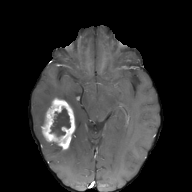}}
    \end{minipage}
    \begin{minipage}{0.2\linewidth}
        \centerline{\includegraphics[width=\textwidth]{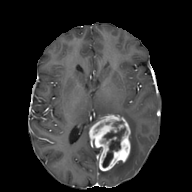}}
    \end{minipage}

    \begin{minipage}{0.12\linewidth}
        \centerline{TSEM}
    \end{minipage}
    \begin{minipage}{0.2\linewidth}
        \centerline{\includegraphics[width=\textwidth]{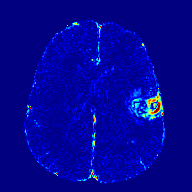}}
    \end{minipage}
    \begin{minipage}{0.2\linewidth}
        \centerline{\includegraphics[width=\textwidth]{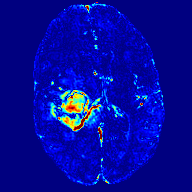}}
    \end{minipage}
    \begin{minipage}{0.2\linewidth}
        \centerline{\includegraphics[width=\textwidth]{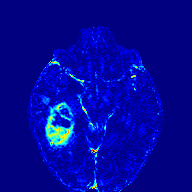}}
    \end{minipage}
    \begin{minipage}{0.2\linewidth}
        \centerline{\includegraphics[width=\textwidth]{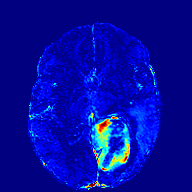}}
    \end{minipage}

    \begin{minipage}{0.12\linewidth}
        \centerline{ResViT}
    \end{minipage}
    \begin{minipage}{0.2\linewidth}
        \centerline{\includegraphics[width=\textwidth]{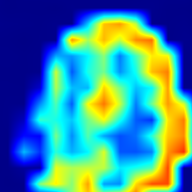}}
    \end{minipage}
    \begin{minipage}{0.2\linewidth}
        \centerline{\includegraphics[width=\textwidth]{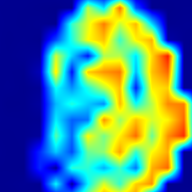}}
    \end{minipage}
    \begin{minipage}{0.2\linewidth}
        \centerline{\includegraphics[width=\textwidth]{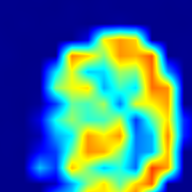}}
    \end{minipage}
    \begin{minipage}{0.2\linewidth}
        \centerline{\includegraphics[width=\textwidth]{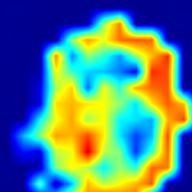}}
    \end{minipage}

    \vspace{5pt}

    \begin{minipage}{0.12\linewidth}
        \centerline{}
    \end{minipage}
    \begin{minipage}{0.2\linewidth}
        \centerline{Case 1}
    \end{minipage}
    \begin{minipage}{0.2\linewidth}
        \centerline{Case 2}
    \end{minipage}
    \begin{minipage}{0.2\linewidth}
        \centerline{Case 3}
    \end{minipage}
    \begin{minipage}{0.2\linewidth}
        \centerline{Case 4}
    \end{minipage}
	
	\caption{Examples of the proposed TSEM and the attention maps extracted by ResViT~\cite{dalmaz2022resvit} when generating T1Gd by given with T1, T2, and Flair.} \label{fig:tsem}
\end{figure}

\begin{table}[t]
    \centering
    \caption{Results of PSNR for regions highlighted or not highlighted by TSEM.\label{tab:tsem}}
    \setlength{\tabcolsep}{3pt}
    \begin{tabular}{cccc}
        \hline\hline
        Number of inputs & $\text{TSEM}>99\%$ & $\text{TSEM}<99\%$ & Total \\
        \hline
        1 & 18.0$\pm$3.2 & 28.3$\pm$2.4 & 27.8$\pm$2.4 \\
        2 & 18.8$\pm$3.7 & 28.8$\pm$2.4 & 28.3$\pm$2.4 \\
        3 & 19.5$\pm$4.0 & 29.3$\pm$2.5 & 28.8$\pm$2.6 \\
        
        \hline\hline
    \end{tabular}
\end{table}

\subsubsection{TSEM vs. Attention Map.}
Figure~\ref{fig:tsem} shows the proposed TSEM and the attention maps extracted by ResViT~\cite{dalmaz2022resvit}.
As shown in Fig.~\ref{fig:tsem}, TSEM has a higher resolution than the attention maps and can highlight the tumor area which is hard to be synthesized by the networks.
Table~\ref{tab:tsem} reports the results of PSNR for regions highlighted or not highlighted by TSEM with a threshold of the 99th percentile.
To assist the synthesis models deploying in clinical settings, TSEM can be used as an attention and uncertainty map to remind clinicians of the possible unreliable synthesized area.

\section{Conclusion}
In this work, we introduce an explainable network for multi-to-one synthesis with extensive experiments and interpretability visualization.
Experimental results based on BraTS2021 demonstrate the superiority of our approach compared with the state-of-the-art methods.
And we will explore the proposed method in assisting downstream applications for multi-sequence analysis in future works.

%
%
%
\bibliographystyle{splncs04}
\bibliography{ref}

\begin{thebibliography}{10}
\providecommand{\url}[1]{\texttt{#1}}
\providecommand{\urlprefix}{URL }
\providecommand{\doi}[1]{https://doi.org/#1}

\bibitem{abnar2020quantifying}
Abnar, S., Zuidema, W.: Quantifying attention flow in transformers. arXiv
  preprint arXiv:2005.00928  (2020)

\bibitem{armanious2020medgan}
Armanious, K., Jiang, C., Fischer, M., K{\"u}stner, T., Hepp, T., Nikolaou, K.,
  Gatidis, S., Yang, B.: Medgan: Medical image translation using gans.
  Computerized medical imaging and graphics  \textbf{79},  101684 (2020)

\bibitem{baid2021rsna}
Baid, U., Ghodasara, S., Mohan, S., Bilello, M., Calabrese, E., Colak, E.,
  Farahani, K., Kalpathy-Cramer, J., Kitamura, F.C., Pati, S., et~al.: The
  rsna-asnr-miccai brats 2021 benchmark on brain tumor segmentation and
  radiogenomic classification. arXiv preprint arXiv:2107.02314  (2021)

\bibitem{bakas2017advancing}
Bakas, S., Akbari, H., Sotiras, A., Bilello, M., Rozycki, M., Kirby, J.S.,
  Freymann, J.B., Farahani, K., Davatzikos, C.: Advancing the cancer genome
  atlas glioma mri collections with expert segmentation labels and radiomic
  features. Scientific data  \textbf{4}(1),  1--13 (2017)

\bibitem{chartsias2017multimodal}
Chartsias, A., Joyce, T., Giuffrida, M.V., Tsaftaris, S.A.: Multimodal mr
  synthesis via modality-invariant latent representation. IEEE transactions on
  medical imaging  \textbf{37}(3),  803--814 (2017)

\bibitem{chefer2021transformer}
Chefer, H., Gur, S., Wolf, L.: Transformer interpretability beyond attention
  visualization. In: Proceedings of the IEEE/CVF Conference on Computer Vision
  and Pattern Recognition. pp. 782--791 (2021)

\bibitem{chen2013clinical}
Chen, J.H., Su, M.Y.: Clinical application of magnetic resonance imaging in
  management of breast cancer patients receiving neoadjuvant chemotherapy.
  BioMed research international  \textbf{2013} (2013)

\bibitem{choi2018stargan}
Choi, Y., Choi, M., Kim, M., Ha, J.W., Kim, S., Choo, J.: Stargan: Unified
  generative adversarial networks for multi-domain image-to-image translation.
  In: Proceedings of the IEEE conference on computer vision and pattern
  recognition. pp. 8789--8797 (2018)

\bibitem{dalmaz2022resvit}
Dalmaz, O., Yurt, M., {\c{C}}ukur, T.: Resvit: residual vision transformers for
  multimodal medical image synthesis. IEEE Transactions on Medical Imaging
  \textbf{41}(10),  2598--2614 (2022)

\bibitem{han2023synthesis}
Han, L., Tan, T., Zhang, T., Huang, Y., Wang, X., Gao, Y., Teuwen, J., Mann,
  R.: Synthesis-based imaging-differentiation representation learning for
  multi-sequence 3d/4d mri. arXiv preprint arXiv:2302.00517  (2023)

\bibitem{huang2018multimodal}
Huang, X., Liu, M.Y., Belongie, S., Kautz, J.: Multimodal unsupervised
  image-to-image translation. In: Proceedings of the European conference on
  computer vision (ECCV). pp. 172--189 (2018)

\bibitem{isola2017image}
Isola, P., Zhu, J.Y., Zhou, T., Efros, A.A.: Image-to-image translation with
  conditional adversarial networks. In: Proceedings of the IEEE conference on
  computer vision and pattern recognition. pp. 1125--1134 (2017)

\bibitem{jung2021conditional}
Jung, E., Luna, M., Park, S.H.: Conditional gan with an attention-based
  generator and a 3d discriminator for 3d medical image generation. In: Medical
  Image Computing and Computer Assisted Intervention--MICCAI 2021: 24th
  International Conference, Strasbourg, France, September 27--October 1, 2021,
  Proceedings, Part VI 24. pp. 318--328. Springer (2021)

\bibitem{li2019diamondgan}
Li, H., Paetzold, J.C., Sekuboyina, A., Kofler, F., Zhang, J., Kirschke, J.S.,
  Wiestler, B., Menze, B.: Diamondgan: unified multi-modal generative
  adversarial networks for mri sequences synthesis. In: Medical Image Computing
  and Computer Assisted Intervention--MICCAI 2019: 22nd International
  Conference, Shenzhen, China, October 13--17, 2019, Proceedings, Part IV 22.
  pp. 795--803. Springer (2019)

\bibitem{li2022virtual}
Li, W., Xiao, H., Li, T., Ren, G., Lam, S., Teng, X., Liu, C., Zhang, J., Lee,
  F.K.h., Au, K.h., et~al.: Virtual contrast-enhanced magnetic resonance images
  synthesis for patients with nasopharyngeal carcinoma using
  multimodality-guided synergistic neural network. International Journal of
  Radiation Oncology* Biology* Physics  \textbf{112}(4),  1033--1044 (2022)

\bibitem{mann2019breast}
Mann, R.M., Cho, N., Moy, L.: Breast mri: state of the art. Radiology
  \textbf{292}(3),  520--536 (2019)

\bibitem{menze2014multimodal}
Menze, B.H., Jakab, A., Bauer, S., Kalpathy-Cramer, J., Farahani, K., Kirby,
  J., Burren, Y., Porz, N., Slotboom, J., Wiest, R., et~al.: The multimodal
  brain tumor image segmentation benchmark (brats). IEEE transactions on
  medical imaging  \textbf{34}(10),  1993--2024 (2014)

\bibitem{sharma2019missing}
Sharma, A., Hamarneh, G.: Missing mri pulse sequence synthesis using
  multi-modal generative adversarial network. IEEE transactions on medical
  imaging  \textbf{39}(4),  1170--1183 (2019)

\bibitem{uzunova2020memory}
Uzunova, H., Ehrhardt, J., Handels, H.: Memory-efficient gan-based domain
  translation of high resolution 3d medical images. Computerized Medical
  Imaging and Graphics  \textbf{86},  101801 (2020)

\bibitem{woo2018cbam}
Woo, S., Park, J., Lee, J.Y., Kweon, I.S.: Cbam: Convolutional block attention
  module. In: Proceedings of the European conference on computer vision (ECCV).
  pp. 3--19 (2018)

\bibitem{zhang2018unreasonable}
Zhang, R., Isola, P., Efros, A.A., Shechtman, E., Wang, O.: The unreasonable
  effectiveness of deep features as a perceptual metric. In: Proceedings of the
  IEEE conference on computer vision and pattern recognition. pp. 586--595
  (2018)

\bibitem{zhang2023important}
Zhang, T., Tan, T., Han, L., Wang, X., Gao, Y., Teuwen, J., Beets-Tan, R.,
  Mann, R.: Important-net: Integrated mri multi-parameter reinforcement fusion
  generator with attention network for synthesizing absent data. arXiv preprint
  arXiv:2302.01788  (2023)

\bibitem{zhou2020hi}
Zhou, T., Fu, H., Chen, G., Shen, J., Shao, L.: Hi-net: hybrid-fusion network
  for multi-modal mr image synthesis. IEEE transactions on medical imaging
  \textbf{39}(9),  2772--2781 (2020)

\end{thebibliography}

\end{document}